\documentclass[a4paper,twoside]{article}
\usepackage[authoryear]{natbib}
\bibpunct{(}{)}{;}{a}{}{,}
\usepackage{graphicx}
\date{} 
\newcommand{\kms}{\ensuremath{\hbox{km}\cdot \hbox{s}^{-1}}}

\newcommand{\f}{\frac}

\newcommand{\bfr}{{\bf r}}
\newcommand{\bfR}{{\bf R}}

\newcommand{\bfv}{{\bf v}}
\newcommand{\bfV}{{\bf V}}

\newcommand{\calM}{{\cal M}}

\newcommand{\bc}{\begin{center}}
\newcommand{\be}{\begin{equation}}
\newcommand{\ee}{\end{equation}}
\newcommand{\ec}{\end{center}}

\newcommand{\spose}[1]{\hbox to 0pt{#1\hss}}
\newcommand{\lta}{\mathrel{\spose{\lower 3pt\hbox{$\mathchar"218$}}
 \raise 2.0pt\hbox{$\mathchar"13C$}}}
\newcommand{\gta}{\mathrel{\spose{\lower 3pt\hbox{$\mathchar"218$}}
 \raise 2.0pt\hbox{$\mathchar"13E$}}}

\title{\large\bf\flushleft Cosmology Under Milne's Shadow}
\author{\parbox{\textwidth}{\flushleft
\vspace{-0.5cm}
{\it Micha{\l} J.\ Chodorowski }\\
\vspace{0.4cm}
{\small Copernicus Astronomical Center, Bartycka 18, 00--716 
Warsaw, Poland}\\
{\small (E-mail: michal@camk.edu.pl)}}}
\begin{document}
\twocolumn[
\begin{changemargin}{.8cm}{.5cm}
\begin{minipage}{.9\textwidth}
\vspace{-1cm}
\maketitle
%
%
\small{\bf Abstract:} Based on the magnitude--redshift diagram for the
sample of supernovae Ia analysed by Perlmutter et al.\ (1999), Davis
\& Lineweaver rule out the special relativistic interpretation of
cosmological redshifts at a confidence level of 23$\sigma$. Here, we
critically reassess this result. Special relativity is known to
describe the dynamics of an empty universe, by means of the Milne
kinematic model. Applying only special-relativistic concepts, we
derive the angular diameter distance and the luminosity distance in
the Milne model. In particular, in this model we do not use the
underlying metric in its Robertson-Walker form, so our exposition is
useful for readers without any knowledge of general relativity. We do
however, explicitly use the special-relativistic Doppler formula for
redshift. We apply the derived luminosity distance to the
magnitude--redshift diagram for supernovae Ia of Perlmutter et al.\
(1999) and show that special relativity fits the data much better than
that claimed by Davis \& Lineweaver. Specifically, {\em using these
data alone\/}, the Milne model is ruled out only at a 2$\sigma$
level. Although not a viable cosmological model, in the context of
current research on supernovae Ia it remains a useful reference model
when comparing predictions of various cosmological models.


\medskip
\medskip
\end{minipage}
\end{changemargin}
]
\small

\section{Introduction}
\label{sec:intro}
In a recent paper, \citet{tamara} attempt\-ed to clarify several
common misconceptions about the expansion of the universe. In
particular, they convincingly pointed out that uniform expansion of an
infinite universe implies that very distant galaxies recede from us
with superluminal (faster-than-light) recession velocities. Moreover,
we can observe such galaxies. This does not violate special relativity
(SR), because their velocities are not measured in any observer's
inertial frame. They are measured in the so-called reference frame of
Fundamental Observers, for which the universe looks homogeneous and
isotropic.

Unfortunately, Davis \& Lineweaver not only clarified some
misconceptions; they also created a new misunderstanding. They claimed
to ``observationally rule out the SR Doppler interpretation of
cosmological redshifts at a confidence level of 23$\sigma$.'' (The
special-relativistic interpretation of redshift is kinematic, i.e.,
the Doppler effect.) As we will explain later, Davis \& Lineweaver did
not apply SR properly. Specifically, they did not consistently use the
definition of the luminosity distance, $D_L$. SR is known to describe
the dynamics of an empty universe \citep{peacock,longair}. Their error
led Davis \& Lineweaver to an expression for the luminosity distance
as a function of redshift that was entirely different from that for an
empty universe.

In the framework of general relativity, the calculation of $D_L$ for
an empty universe is straightforward. However, here we will present
also an alternative approach, based entirely on SR. Namely, we will
derive $D_L$ applying the kinematic cosmological model of
\citet{milne}.  This approach is useful for readers without any
knowledge of general relativity. It elucidates the meaning of time and
kinematics in cosmology. The Milne model offers also an interesting
insight into all Friedman-Robertson-Walker (FRW) cosmological models.

The paper is organized as follows. We begin in Sec.~\ref{sec:ang} by
deriving the angular diameter distance for an empty universe using the
FRW framework. Next, we present the corresponding derivation in the
Milne model. In Sec.~\ref{sec:lum} we derive the luminosity distance
in the Milne model and compare it to the angular diameter
distance. Finally, in Sec.~\ref{sec:app} we present the resulting
magnitude--redshift diagram for supernovae Ia. We summarize in
Sec.~\ref{sec:summ}.

\section{Angular diameter distance}
\label{sec:ang}
For comparison, let us first recall the derivation of the angular
diameter distance for an empty universe in the FRW framework. The
metric of a homogeneous and isotropic universe is given by the
Robertson--Walker line element:
\begin{equation}
c^2 ds^2 = c^2 dt^2 - a^2(t)[dx^2 + R_o^2 S^2(x/R_o) d\Omega] .
\label{eq:RW}
\end{equation}
Here, 
\be
d\Omega = d\theta^2 + \sin^2\theta d\phi^2 ,
\label{eq:dOmega}
\end{equation}
and $R_o^{-2}$ is the (present) curvature of the universe.
The function $S(x)$ equals $\sin(x)$, $x$, and $\sinh(x)$ for a
closed, flat, and open universe, respectively. The function $a(t)$ is
called a scale factor and relates the physical, or proper, coordinates
of a galaxy, $\bfr$, to its fixed or comoving coordinates,
$\mathbf{x}$: $\bfr = a \mathbf{x}$. This function accounts for
the expansion of the universe; its detailed time dependence is
determined by the Friedman equations. We normalize $a$ so that at the
present time, $a(t_o) = 1$. For an open universe, the Friedman
equations yield
\be
R_o = \frac{c H_o^{-1}}{(1 - \Omega_o)^{1/2}} ,
\label{eq:curv}
\ee where $H_o$ is the present value of the Hubble constant and
$\Omega_o$ is the present value of the total (energy) density in the
universe, in units of the so-called critical (energy) density. 

Let us work out the angular size of an object of proper length $\Delta
y$, perpendicular to the radial coordinate at redshift $z$. The
relevant spatial component of the
metric~(\ref{eq:RW})--(\ref{eq:dOmega}) is the term in $d \theta$. The
proper length $\Delta y$ of an object at redshift $z$, corresponding
to scale factor $a(t_e)$, for an open universe is \be \Delta y =
a(t_e) R_o \sinh(x/R_o) \Delta \theta = D_A \Delta \theta ,
\label{eq:perp}
\ee 
where we have introduced the angular diameter distance $D_A =
a(t_e) R_o \sinh(x/R_o)$. Here, $t_e$ is the time of emission of
photons. Since $a(t_e)^{-1} = 1 + z$, we have 
\be
D_A = R_o \sinh(x/R_o)/(1 + z). 
\label{eq:D_A_FRW}
\ee
For an {\em empty\/} universe, $\Omega_o = 0$, hence $R_o = c
H_o^{-1}$. Moreover, then $a(t) = H_o t$ and the equations of null
radial geodesics are easy to integrate. The result is $x = c H_o^{-1}
\ln(1+z)$. Substituting this into equation~(\ref{eq:D_A_FRW}) and
using the definition of the hyperbolic sine, we obtain
\be
D_A(z) = c H_o^{-1} \frac{z(1 + z/2)}{(1 + z)^2} .
\label{eq:D_A_FRW_fin}
\ee
This is the angular diameter distance for an empty universe, derived
in the FRW framework. 

In the Milne model, the cosmic arena of physical events is the
pre-existing Minkowski spacetime. In the origin of the coordinate
system, $O$, at time $t = 0$ an `explosion' takes place, sending
radially fundamental observers with constant velocities in the range
of speeds $(0,c)$. The fundamental observer with velocity $v$, $F_v$,
carries a rigid rod of length $\Delta y$, oriented perpendicularly to
the line of sight of the observer at $O$. At time $t_e$ this rod emits
photons. At the photons' arrival time at $O$, $t_o$, the rod subtends
at $O$ an angle
\be
\Delta \phi = \Delta y / r_e \,, 
\label{eq:angle}
\ee
where $r_e$ is the distance from $O$ to $F_v$ at the time of {\em
emission\/} of the photons, $t_e$. We have $t_o = t_e + t_t$, where
$t_t$ is the travel time of the photons. Since
\be
t_e = \frac{r_e}{v} 
\label{eq:t_e} 
\ee
and
\be
t_t = \frac{r_e}{c}, 
\label{eq:t_t}
\ee
we obtain 
\be
r_e = c t_o \frac{\beta}{1 + \beta} , 
\label{eq:r_e}
\ee
where $\beta = v/c$. The special-relativistic formula for the Doppler
effect is 
\be
1 + z = \left(\f{1 + \beta}{1 - \beta}\right)^{1/2} ,
\label{eq:redshift}
\ee
where $z$ is the photons' redshift; hence,
\be
\beta = \frac{(1+z)^2 - 1}{(1+z)^2 + 1}  .
\label{eq:beta}
\ee
Using equation~(\ref{eq:beta}) in~(\ref{eq:r_e}) yields
\be
r_e = c t_o \frac{z(1 + z/2)}{(1 + z)^2} .
\ee
Since in the Milne model, for any time $t$ and for any fundamental
observer $F_v$, $r = v\, t$, the observer at $O$ observes the Hubble
flow: $v = H r$, where the Hubble constant is $H = t^{-1}$. Hence,
$t_o = H_o^{-1}$. The angular diameter distance is defined via the
equation $\Delta y = D_A \Delta \phi$. Therefore, using
equation~(\ref{eq:angle}) we obtain finally
\be
D_A(z) = r_e = c H_o^{-1} \frac{z(1 + z/2)}{(1 + z)^2} .
\label{eq:D_A}
\ee

In the above derivation we have applied special relativity a number
of times. First, writing Eq.~(\ref{eq:t_t}) we have followed its
central assumption, that the velocity of light is always $c$,
regardless the relative motion of the emitter and the
observer. Secondly, we have applied the special-relativistic,
kinematic interpretation of redshift -- i.e., the Doppler effect --
and used formula~(\ref{eq:redshift}) for it. Finally, writing
Eq.~(\ref{eq:angle}) we have assumed that the geometry of space is
Euclidean.

Eq.~(\ref{eq:D_A}) exactly coincides with
formula~(\ref{eq:D_A_FRW_fin}) for $D_A$ for an empty universe, the
latter derived from the metric in its FRW form. Our second derivation,
which was purely special-re\-la\-ti\-vi\-stic, employed other (i.e.,
conventional Minkowskian) definitions of distance and time. Still, we
arrived at the same formula for $D_A$. This is so because this formula
relates direct observables: proper (i.e., rest-frame) size of an
object to its angular size and redshift. Regardless of what are the
definitions of coordinates in a given coordinate system, their
consistent application should lead to the same result in terms of
observables!

\section{Luminosity distance}
\label{sec:lum}
The relation between the luminosity distance, $D_L$, and the angular
diameter distance is 
\be
D_L = (1 + z)^2 D_A .
\label{eq:D_L-D_A}
\ee
However, this is true in general relativity; let's check whether this
holds also in the Milne model. Let's place a source of radiation at
the origin of the coordinate system, $O$. We assume again that at time
$t = 0$ at $O$ an `explosion' takes place, sending radially
fundamental observers with constant velocities in the range of
$(0,c)$. At time $t_e$ the source emits photons, which at time $t_o$
reach a fundamental observer moving with velocity $v$, such that
\be
v t_o = c (t_o - t_e) .
\label{eq:v}
\ee
If the source emits continuously photons with constant bolometric
luminosity $L$, the observer at $r_o = v t_o$ receives a flux of
radiation with bolometric intensity 
\be
f = \frac{L}{4\pi r_o^2 (1 + z)^2} .
\label{eq:flux}
\ee
The factor $(1 + z)^2$ in the denominator is due to the Doppler
effect. Specifically, one factor $1 + z$ is due to the fact that the
wavelength, and so the energy, of the observed photons is
redshifted. The second factor $1 + z$ is due to the fact that photons,
emitted in the time interval $\Delta t_e$, arrive to the observer in
the time interval $t_o = (1 + z) \Delta t_e$.

The luminosity distance is defined by the equation $f = L/(4\pi
D_L^2)$, hence 
\be
D_L = (1 + z) r_o = (1 + z) c \beta t_o .
\label{eq:D_L}
\ee 
Time $t_o$ is the time of observation indicated by the clock at the
source, but the fundamental observer is moving with respect to $O$, so
according to SR, his clock {\em delays\/} compared to that at $O$. At
the moment of observation, his clock shows time $\tau_o = t_o/\gamma$,
where $\gamma = (1 - \beta^2)^{-1/2}$. Therefore, we have 
\be 
D_L = (1 + z) c \tau_o \beta \gamma .
\label{eq:D_L2}
\ee 
Relative to the observer, however, this is the source that is
moving. Using his clock, the observer deduces that since the Big-Bang,
the source has moved off to the distance $r_o' = v \tau_o$. Because
the source is moving, the distance $r_o'$ is length-contracted
relative to its rest-frame value, $r_o$: $r_o' = r_o/\gamma$. Hence,
the observer will agree that $r_o = \gamma v \tau_o = c \tau_o \beta
\gamma$, what yields equation~(\ref{eq:D_L2}).

Next, we have
\be
\beta \gamma = \frac{\beta}{(1-\beta)^{1/2} (1+\beta)^{1/2}} = (1 + z)
\frac{\beta}{1+\beta} ,
\label{eq:betagamma}
\ee
where in the last equality we have used the SR formula for redshift,
Eq.~(\ref{eq:redshift}). Combined with Eq.~(\ref{eq:beta}),
Eqs.~(\ref{eq:D_L2})--(\ref{eq:betagamma}) yield
\be
D_L = (1 + z)^2 c \tau_o \frac{z(1 + z/2)}{(1 + z)^2} = 
c H_o^{-1} z(1 + z/2) .
\label{eq:D_L_fin}
\ee
Comparing with Eq.~(\ref{eq:D_A}) we see that indeed $D_L = (1 + z)^2
D_A$, in accordance with general relativity. 

To derive the angular diameter distance, in Sec.~\ref{sec:ang} we
have used the observer's rest-frame. To derive the luminosity
distance, however, in the present section we have switched to the
source's rest-frame. We have done so for simplicity of the resulting
calculations. In particular, only in the latter frame is the radiation
of the source isotropic, and one can apply simple
equation~(\ref{eq:flux}) for the observed flux.

Deriving the two distances, we have placed either the observer or the
source at a special position: at the center of expansion. Are then our
results general? Yes: although in the Milne model this center does
indeed exist, every fundamental observer considers himself to be at
the center of expansion! This can be easily seen in the
non-relativistic regime. According to the Galilean transformation of
velocities, the velocity of any observer $O''$ relative to another
observer $O'$ is $\bfv' = \bfv -\bfV_{\rm rel}$, where $\bfv$ and
$\bfV_{\rm rel}$ are respectively the velocity of $O''$ relative to
$O$ and the velocity of $O'$ relative to $O$. But by the construction
of the model, the observer at $O$ observes the Hubble flow, so $\bfv =
H_o \bfr$ and $\bfV_{\rm rel} = H_o \bfR$, where $\bfr$ denotes the
position of $O''$ relative to $O$ and $\bfR$ denotes the position of
$O'$ relative to $O$. Hence, $\bfv' = H_o (\bfr - \bfR) = H_o \bfr'$,
where $\bfr'$ is the position of $O''$ relative to $O'$.  Thus, the
observer $O'$ observes an isotropic Hubble flow around him, so he is
apparently at the center of expansion. The point is that this result
holds also for relativistic velocities \citep{milne,rindler}. Strictly
speaking, in the Hubble law the Hubble constant is an inverse of the
local proper time of the observer at $O'$. This is why in
Eq.~(\ref{eq:D_L_fin}) we have identified $H_o$ with $\tau_o^{-1}$.

It is rather surprising why Milne, who insisted so much on an
observables-oriented approach to cosmology, did not derive himself an
explicit formula for the luminosity distance as a function of
redshift. We will see, however, that he was close to it. The Appendix
to his classical paper on kinematic relativity \citep{milne} bears the
title ``The apparent brightness of a receding nebula''. Its final
formula~(16) describes ``the {\em total\/} light received by A (the
observer) on his own photographic plate'' in a fairly complex,
integral form. This formula involves implicitly the luminosity
distance, but in terms of the recession velocity of the nebula rather
than its redshift. However, while the redshift of a distant nebula is
a direct observable, its velocity is not. Extracting the luminosity
distance from the formula and using the SR relation between velocity
and redshift, Eq.~(\ref{eq:redshift}), we rederive the formula for the
luminosity distance given by our Eq.~(\ref{eq:D_L_fin}).

\section{Apparent magnitude -- redshift relation}
\label{sec:app}
\begin{figure}[h]
\begin{center}
\includegraphics[angle=0,scale=0.45]{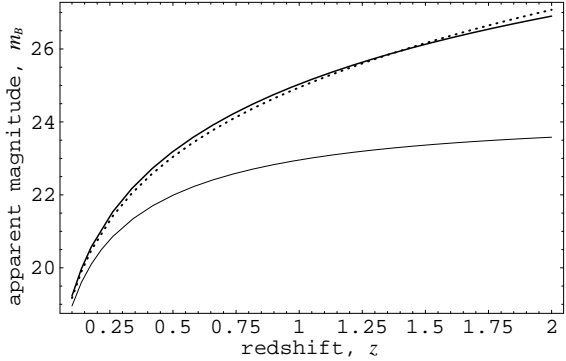}
\caption{\small 
The magnitude--redshift diagram for supernovae Ia. The 
thick solid line shows the prediction of the currently favored
cosmological model, $\Omega_m = 0.28$ and $\Omega_\Lambda = 0.72$.
The thin solid line shows the prediction of special relativity
according to Davis \& Lineweaver. The dotted line shows the {\em
correct\/} prediction of special relativity.}\label{fig:hubble}
\end{center}
\end{figure}

The apparent bolometric magnitude, $m_B$, of a standard candle located
at redshift $z$ is related to its absolute bolometric magnitude, $M$,
by the equation
\be
m_B = 5 \log_{10} D_L + 25 + M .
\label{eq:hubble}
\ee
Here, $D_L$ is the luminosity distance, expressed in me\-gaparsecs
(Mpc). Supernovae Ia turned out to be very good standard candles in
the universe \citep[see, for example,][]{perl}. \citet{p97} cast the
above equation to the form
\be
m_B = 5 \log_{10} (H_o D_L) + \calM ,
\label{eq:p97}
\ee
where
\be
\calM = M - 5 \log_{10} H_o + 25 
\label{eq:calM}
\ee
is the magnitude zero-point, and the Hubble constant is expressed in
$\kms \cdot \hbox{Mpc}^{-1}$ (Eq.~(1)--(2) of Perlmutter et al.\ 1997). 
We prefer to rewrite Eq.~(\ref{eq:p97}) to the form
\be
m_B = 5 \log_{10} \left(\frac{D_L}{c H_o^{-1}}\right) +
\widetilde\calM ,
\label{eq:myform}
\ee
where
\be
\widetilde\calM = \calM + 5 \log_{10}c = \calM + 25 + 
5 \log_{10} 2.998 ;
\label{eq:myM}
\ee
the argument of logarithm in Eq.~(\ref{eq:myform}) is explicitly
dimensionless. Introducing Eq.~(\ref{eq:D_L_fin}) in~(\ref{eq:myform})
yields
\be
m_B = 5 \log_{10} \left[z(1 + z/2)\right] + \widetilde\calM .
\label{eq:milne}
\ee
This is the magnitude--redshift relation in the Milne model. It
coincides exactly with the corresponding relation for an empty
universe. In Fig.~\ref{fig:hubble} it is shown as a dotted line. For
the magnitude zero-point we adopt the value $\calM = -3.32$,
calculated in \citet{p97} and used also in \citet{p99}.

To derive the luminosity distance $D_L = (1 + z)D$,
Davis \& Lineweaver used the Hubble law:
\be
D = c H^{-1} \beta .
\label{eq:D}
\ee
Comparing with Eq.~(\ref{eq:D_L}) we see that this leads to the
correct expression for $D_L$ provided we identify $H$ with
$H_o^{(s)}$, i.e. the value of the Hubble constant at the source and
at the time of {\em observation} of photons. However, they identified
$H$ with $H_e^{(s)}$, i.e. the Hubble constant at the source but at
the time of {\em emission} of photons:
\be
D^{(DL)} = c \left[H_e^{(s)}\right]^{-1} \beta = c \beta t_e.  
\label{eq:D^DL}
\ee 
Combined with the SR expression for $\beta$ as a function of redshift,
Eq.~(\ref{eq:beta}), the above equation yields
\be
D^{(DL)} = \frac{c}{H_e^{(s)}} \, \frac{(1+z)^2 - 1}{(1+z)^2 + 1}
.
\label{eq:D^DL2}
\ee
\citep[Eq.~(10) of][]{tamara}. Next, Davis \& Lineweaver applied 
the equality
\be
H_e^{(s)} = (1 + z) H_o^{(o)}, 
\label{eq:He_Ho}
\ee 
correct in case of an empty universe. Let's check whe\-ther this holds
also in the Milne model. First, $H_e^{(s)} = t_e^{-1}$ and $H_o^{(o)}
= \tau_o^{-1}$, where $t$ and $\tau$ are the time measured (by
Fundamental Observers) respectively at the source and at the observing
point. Hence, 
\be 
\frac{H_e^{(s)}}{H_o^{(o)}} = \frac{\tau_o}{t_e} = \gamma^{-1}
{p}^{-1} ,
\label{eq:H_ratio}
\ee
where ${p} \equiv t_e/t_o$. From Eq.~(\ref{eq:v}) we have 
\be
\beta = 1 - {p} , 
\label{eq:beta_M}
\ee
so
\be
\gamma^{-1} = (1 - \beta^2)^{1/2} = \left[{p}(2-{p})\right]^{1/2} .
\label{eq:gamma}
\ee
Used in Eq.~(\ref{eq:H_ratio}), the above equation yields
\be
\frac{H_e^{(s)}}{H_o^{(o)}} = \left(\frac{2-{p}}{{p}}\right)^{1/2} .
\label{eq:H_ratio2}
\ee
In turn, Eq.~(\ref{eq:beta_M}), used in Eq.~(\ref{eq:redshift}),
yields 
\be
1 + z = \left(\frac{2-{p}}{{p}}\right)^{1/2} .
\label{eq:1+z}
\ee By inspection of Eqs.~(\ref{eq:H_ratio2})--(\ref{eq:1+z}) we see
that they indeed imply Eq.~(\ref{eq:He_Ho}). Once again, we have found
that the Milne model exactly describes the dynamics of an empty
universe. 

Comparing equation~(\ref{eq:D^DL}) with~(\ref{eq:D_L}) we see that
equation~(\ref{eq:D^DL}) leads to an expression for the luminosity
distance which {\em underestimates\/} the correct one by the factor of
$t_o/t_e$. We have $t_o/t_e = (t_o/\tau_o)\cdot(\tau_o/t_e) = \gamma
(1 + z) = [(1+z)^2 + 1]/2$. Using Eq.~(\ref{eq:He_Ho}) in
Eq.~(\ref{eq:D^DL2}) we obtain
\be D^{(DL)} = (1 + z)^{-1}
\frac{c}{H_o^{(o)}} \, \frac{(1+z)^2 - 1}{(1+z)^2 + 1} ,
\label{eq:D^(DL)}
\ee
therefore
\be
D_L^{(DL)} = (1 + z) D^{(DL)} = 
\frac{c}{H_o} \, \frac{(1+z)^2 - 1}{(1+z)^2 + 1} .
\label{eq:D_L^(DL)}
\ee 
Here, $H_o$ denotes the value of the Hubble constant at the observing
point at the time of observation and for simplicity we have skipped
the superscript ``$^{(o)}$''. Combined with Eq.~(\ref{eq:myform}),
Eq.~(\ref{eq:D_L^(DL)}) yields the function shown in
Fig.~\ref{fig:hubble} as a thin solid line. Davis \& Lineweaver did
not provide an explicit formula for the magnitude--redshift relation
in SR, but this line closely follows their plotted curve.

The thick solid line in Fig.~\ref{fig:hubble} shows the
magnitude--redshift relation for the currently favored cosmological
model: a flat universe with a nonzero cosmological constant, $\Omega_m
= 0.28$ and $\Omega_\Lambda = 0.72$.\footnote{These are recent
estimates of the cosmological parameters $\Omega_m$ and
$\Omega_\Lambda$ from the magnitude--redshift relation of observed
high-redshift supernovae, combined with other observational
constraints. See, for example, \citet{tonry}.} Useful expressions for
the luminosity distance in this model are given in \citet{mch}. We see
that the thin solid line (prediction of SR according to
Davis \& Lineweaver) is very distant from the thick solid line. Using
the supernovae data of \citet{p99}, Davis \& Lineweaver verified that
the model given by the thin line ``is ruled out at more than
23$\sigma$'' compared with the currently favored model.

On the other hand, the dotted line (correct prediction of SR) follows
the thick solid line much more closely. Given the data by \citet{p99},
how close is the prediction of SR to that of the favored model? The
answer is provided by Perlmutter et al. themselves: their analysis
yielded the constraint $0.8 \Omega_m - 0.6
\Omega_\Lambda = -0.2 \pm 0.1$. An empty universe corresponds to
$\Omega_m = \Omega_\Lambda = 0$, hence for SR, $0.8
\Omega_m - 0.6 \Omega_\Lambda = 0$. We see thus that {\em within two 
standard deviations,\/} the data was consistent with an empty
universe! This is not to defend this model as a viable alternative to
the currently favored model. From our mere existence we know the
universe is not empty. A host of observational evidence consistently
points towards the currently favored model. This is only to say that a
few years ago, the evidence {\em from supernovae data alone\/} for the
accelerated expansion of the universe was not so strong, and the 
assumption of its purely kinematic expansion at low redshifts could
then serve as a reasonable starting approximation.

Since the year 1999 the supernovae Ia data has improved. In
particular, the analysis of \citet{tonry} yielded $\Omega_\Lambda -
1.4 \Omega_m = 0.35 \pm 0.14$. This constituted a modest improvement
over the result of Perlmutter et al., implying the present acceleration
of the universe's expansion to be detected at a $2.5\sigma$
confidence.\footnote{From supernovae alone. Combined with the
constraint of a flat universe, strongly supported by the CMB
observations, the data of \citeauthor{tonry} yielded $\Omega_m = 0.28
\pm 0.05$ and $\Omega_\Lambda = 0.72 \pm 0.05$, implying a currently
accelerating universe at much higher confidence.} A significant
improvement was achieved by discovering and observing supernovae Ia at
$z > 1$ with the Hubble Space Telescope. Fig.~8 of \citet{riess} shows
the resulting joint confidence intervals for ($\Omega_m$,
$\Omega_\Lambda$) from SNe Ia. In this figure, the point $\Omega_m =
\Omega_\Lambda = 0$ lies outside shown confidence contours of
$99.7$\%. \citeauthor{riess} claim that ``with the current sample, the
$4\sigma$ confidence intervals (i.e., $> 99.99$\% confidence) are now
fully contained within the region where $\Omega_\Lambda > 0$.''
Similar results of the analysis of the \citeauthor{riess} sample have
been obtained independently by \citet{wright}.

An empty universe is thus not a viable cosmological model, but remains
a useful reference model when comparing predictions of various
cosmological models. Fig.~7 of \citeauthor{riess} shows the
magnitude--redshift diagram for SN Ia in a residual form: relative to
an empty universe model. Being eternally coasting, this model has a
vanishing deceleration parameter, so it naturally separates
accelerating from decelerating models. Also, it is evident from Fig.~7
of \citeauthor{riess} that the model fits the data much better than an
Einstein--de Sitter universe ($\Omega_m = 1$, $\Omega_\Lambda = 0$).

\section{Summary}
\label{sec:summ}
We have derived the angular diameter distance $D_A$ and the luminosity
distance $D_L$ in the Milne kinematic cosmological model, using only
special-relativistic concepts. In the derivations, the central r\^ole
was played by the special-relativistic Doppler formula for photons'
redshift. We have found that $D_L = (1 + z)^2 D_A$, in accordance with
general relativity. The derived formulae are identical to these
corresponding to an empty universe in the FRW cosmology. We have shown
where Davis \& Lineweaver failed to correctly derive the luminosity
distance. Finally, we have presented the resulting magnitude--redshift
diagram for supernovae Ia. While the prediction of special relativity
according to Davis \& Lineweaver is far away from that for the
currently favored cosmological model, the correct prediction of
special relativity follows the favored model much more closely. Though
not a viable alternative to the currently favored model, the Milne
model has great pedagogical value, elucidating the kinematic aspect of
the universe's expansion. In the context of current research on
supernovae Ia, it remains a useful reference model when comparing
predictions of various cosmological models.

\section*{Acknowledgments}
This research has been supported in part by the Polish State Committee
for Scientific Research grant No.~1 P03D 012 26, allocated for the
period 2004--2007.


\begin{thebibliography}{}

\bibitem[Chodorowski(2005)]{mch} Chodorowski, M.~J. 2005, AmJPh, 
in press
\bibitem[Davis \& Lineweaver(2004)]{tamara} Davis, T.~M., 
Lineweaver, C.~H. 2004, PASA, 21, 97
\bibitem[Longair(2003)]{longair} Longair, M.~S. 2003, Theoretical 
Concepts in Physics, (Cambridge: Cambridge University Press), 2nd ed.,
p.\ 543
\bibitem[Milne(1933)]{milne} Milne, E.~A. 1933, Zeitschrift f\"ur 
Astrophysik, 6, 1
\bibitem[Peacock(1999)]{peacock} Peacock, J.~A. 1999, Cosmological 
Physics, (Cambridge: Cambridge University Press), p.\ 88
\bibitem[Perlmutter et al.(1997)]{p97} Perlmutter, S., et al.\ 1997, 
ApJ, 483, 565
\bibitem[Perlmutter et al.(1999)]{p99} Perlmutter, S., et al.\ 1999, 
ApJ, 517, 565
\bibitem[Perlmutter(2003)]{perl} Perlmutter, S. 2003, Physics Today, 
56, 53
\bibitem[Riess et al.(2004)]{riess} Riess, A.~G., et al.\ 2004, ApJ, 
607, 665
\bibitem[Rindler(1977)]{rindler} Rindler, W. 1977, Essential 
Relativity, (New York, Heidelberg, Berlin: Springer-Verlag), revised 
2nd ed., Sec.\ (9.4)
\bibitem[Tonry et al.(2003)]{tonry} Tonry, J.~L., et al.\ 2003, ApJ, 
594, 1
\bibitem[Wright(2005)]{wright} Wright, E.~L. 2005, \raggedright
\verb+www.astro.ucla.edu/~wright/sne_cosmology.html+

\end{thebibliography}
\end{document}